\documentclass[sigconf]{acmart}
\pdfoutput=1
\usepackage{subfigure, multirow, tabularx}
\usepackage{booktabs}
\usepackage[boxed, ruled,vlined,linesnumbered]{algorithm2e}
\usepackage{enumitem}
\usepackage{flushend}

\usepackage{amsmath, amsfonts, amssymb, amsthm, xspace, color}

\newcommand{\hide}[1]{} 

\newcommand{\benum}[1]{\begin{enumerate}#1\end{enumerate}}
\newcommand{\bitem}[1]{\begin{itemize}#1\end{itemize}}

\newcommand{\etal}{\emph{et~al.}\xspace} 
\newcommand{\ie}{\emph{i.e.}\xspace} 
\newcommand{\eg}{\emph{e.g.}\xspace} 

\newcommand{\Sum}{\sum\limits} 

\def \v {\mathbf{v}}

\def \H {\mathbf{H}}

\def \D {\mathcal{D}}

\def \H {\mathcal{H}}

\def \S {\mathcal{S}}
\def \T {\mathcal{T}}

\newcommand{\dblp}{\textsc{DBLP}\xspace}
\newcommand{\sigp}{\textsc{SP}\xspace}
\newcommand{\kmeans}{\textsc{Spherical-Kmeans}\xspace}

\newcommand{\our}{\textsc{TaxoGen}\xspace}
\newcommand{\hc}{\textsc{HClus}\xspace}
\newcommand{\hlda}{\textsc{HLDA}\xspace}
\newcommand{\hpam}{\textsc{HPAM}\xspace}
\newcommand{\nole}{\textsc{NoLE}\xspace}
\newcommand{\noac}{\textsc{NoAC}\xspace}

\begin{document}

\copyrightyear{2018}
\acmYear{2018}
\setcopyright{acmcopyright}
\acmConference[KDD 2018]{24th ACM SIGKDD International Conference on Knowledge Discovery \& Data Mining}{August 19--23, 2018}{London, United Kingdom}
\acmBooktitle{KDD 2018: 24th ACM SIGKDD International Conference on Knowledge Discovery \& Data Mining, August 19--23, 2018, London, United Kingdom}
\acmPrice{15.00}
\acmDOI{10.1145/3219819.3220064}
\acmISBN{978-1-4503-5552-0/18/08}


\fancyhead{}

\title{TaxoGen: Unsupervised Topic Taxonomy Construction by Adaptive Term Embedding and Clustering}

\author{
  Chao Zhang$^1$,  Fangbo Tao$^2$, Xiusi Chen$^3$, Jiaming Shen$^1$, Meng Jiang$^4$,\\
  Brian Sadler$^5$, Michelle Vanni$^5$, and Jiawei Han$^1$
}
\affiliation{
  \institution{$^1$Dept. of Computer Science, University of Illinois at Urbana-Champaign, Urbana, IL, USA}
  \institution{$^2$Facebook Inc., Menlo Park, CA, USA}
  \institution{$^3$Dept. of Computer Science and Technology, Peking University, Beijing, China}
  \institution{$^4$Dept. of Computer Science and Engineering, University of Notre Dame, Notre Dame, IN, USA}
  \institution{$^5$U.S. Army Research Laboratory, Adelphi, MD, USA}
  \institution{$^{1}$\{czhang82, js2, hanj\}@illinois.edu ~~~ $^2$fangbo.tao@gmail.com ~~~ $^3$xiusi0721@gmail.com} 
  \institution{$^4$mjiang2@nd.edu ~~~ $^{5}$\{brian.m.sadler6.civ, michelle.t.vanni.civ\}@mail.mil}
}

\begin{abstract}

  Taxonomy construction is not only a fundamental task for semantic analysis of
  text corpora, but also an important step for applications such as information
  filtering, recommendation, and Web search.  Existing pattern-based methods
  extract hypernym-hyponym term pairs and then organize these pairs into a
  taxonomy. However, by considering each term as an independent concept node,
  they overlook the topical proximity and the semantic correlations among
  terms.  In this paper, we propose a method for constructing \emph{topic
    taxonomies}, wherein every node represents a conceptual topic and is
  defined as a cluster of semantically coherent concept terms. Our method,
  \our, uses term embeddings and hierarchical clustering to construct a topic
  taxonomy in a recursive fashion. To ensure the quality of the recursive
  process, it consists of: (1) an adaptive spherical clustering module for
  allocating terms to proper levels when splitting a coarse topic into
  fine-grained ones; (2) a local embedding module for learning term embeddings
  that maintain strong discriminative power at different levels of the
  taxonomy. Our experiments on two real datasets demonstrate the effectiveness
  of \our compared with baseline methods.

\end{abstract}

\maketitle

\section{Introduction}
\label{sect:intro}

Automatic taxonomy construction from a text corpus is a fundamental task for
semantic analysis of text data and plays an important role in many
applications. For example, organizing a massive news corpus into a
well-structured taxonomy allows users to quickly navigate to their interested
topics and easily acquire useful information. As another example, many
recommender systems involve items with textual descriptions, and a taxonomy for
these items can help the system better understand user interests to make more
accurate recommendations \cite{Zhang2014TaxonomyDF}.

Existing methods mostly generate a taxonomy wherein each node is a single term
representing an independent concept \cite{Hearst92, LuuKN14}.  They use
pre-defined lexico-syntactic patterns (\eg, A such as B, A is a B) to extract
hypernym-hyponym term pairs, and then organize these pairs into a concept
taxonomy by considering each term as a node.  Although they can achieve high
precision for the extracted hypernym-hyponym pairs, considering each term as an
independent concept node causes three critical problems to the taxonomy: (1)
\emph{low coverage}: Since term correlations are not considered, only the pairs
exactly matching the pre-defined patterns are extracted, which leads to low
coverage of the result taxonomy. (2) \emph{high redundancy}: As one concept can
be expressed in different ways, the taxonomy is highly redundant because many
nodes are just different expressions of the same concept (\eg, `information
retrieval' and `ir').  (3) \emph{limited informativeness}: Representing a node
with a single term provides limited information about the concept and causes
ambiguity.

\begin{figure}[t]
  \centering
  \centerline{\includegraphics[width=1.0\columnwidth]{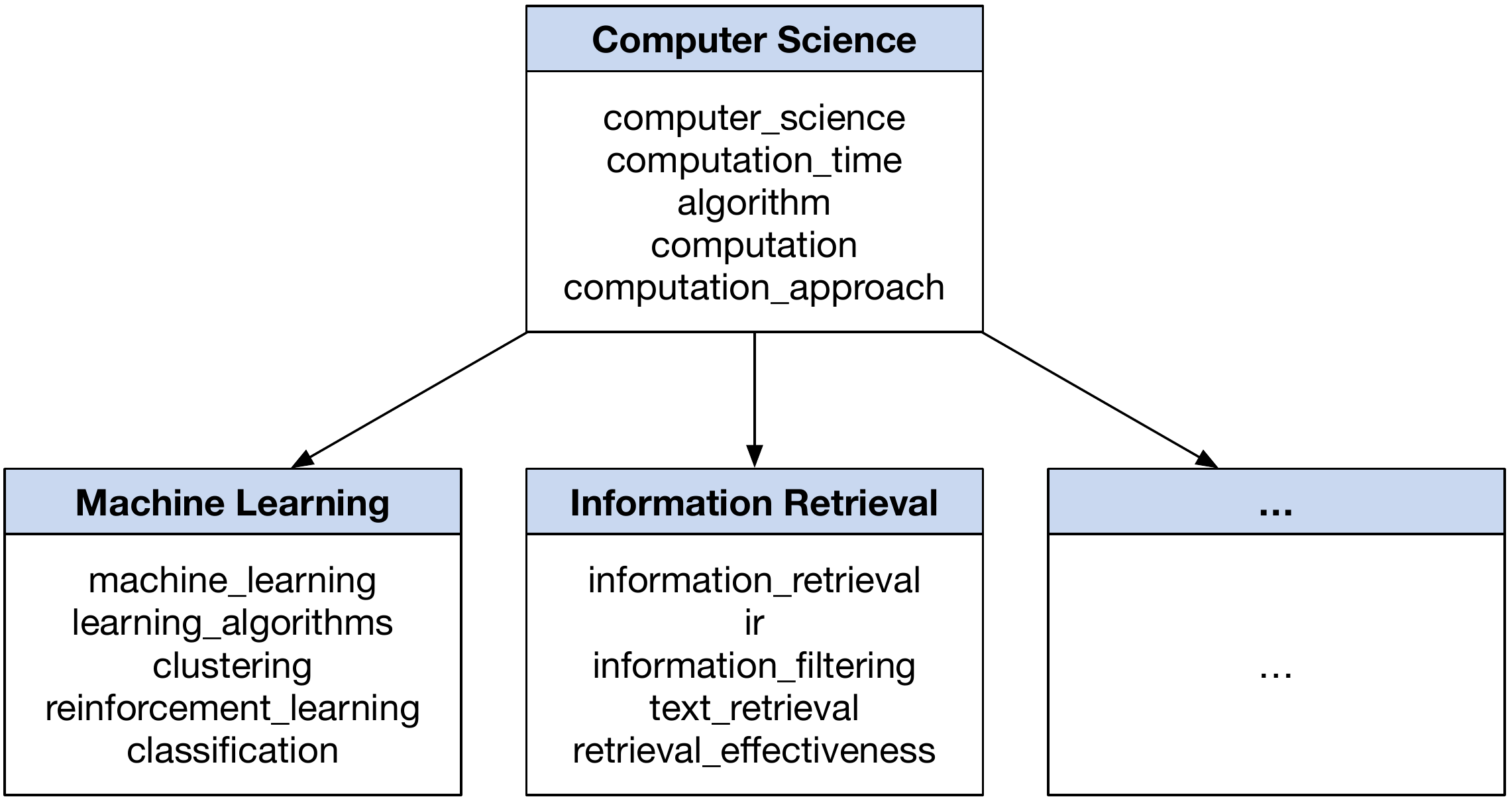}} 
  \caption{An example topic taxonomy. Each node is a cluster of semantically
    coherent concept terms representing a conceptual topic.}
  \label{fig:intro}
\end{figure}

We study the problem of \emph{topic taxonomy construction} from an input text
corpus.  In contrast to term-level taxonomies, each node in our topic taxonomy
represents a conceptual topic, defined as a cluster of semantically coherent
concept terms.  Figure \ref{fig:intro} shows an example.  Given a collection of
computer science research papers, we build a tree-structured hierarchy. The
root node is the general topic `computer science', which is further split into
sub-topics like `machine learning' and `information retrieval'.  For every
topical node, we describe it with multiple concept terms that are semantically
relevant. For instance, for the `information retrieval' node, its associated
terms include not only synonyms of `information retrieval' (\eg, `ir'), but
also different facets of the IR area (\eg, `text retrieval' and `retrieval
effectiveness').

We propose an unsupervised method named \our for constructing topic taxonomies.
It embeds the concept terms into a latent space to capture their semantics, and
uses term embeddings to recursively construct the taxonomy based on
hierarchical clustering.  While the idea of combining term embedding and
hierarchical clustering is intuitive by itself, two key challenges need to be
addressed for building high-quality taxonomies. First, \emph{it is nontrivial
  to determine the proper granularity levels for different concept terms}.
When splitting a coarse topical node into fine-grained ones,  not all the
concept terms should be pushed down to the child level.  For example, when
splitting the computer science topic in Figure \ref{fig:intro}, general terms
like `cs' and `computer science' should remain in the parent instead of being
allocated into any child topics.  Therefore, it is problematic to directly
group parent terms to form child topics, but necessary to allocate different
terms to different levels.  Second, \emph{global embeddings have limited
  discriminative power at lower levels}.  Term embeddings are typically learned
by collecting the context evidence from the corpus, such that terms sharing
similar contexts tend to have close embeddings. However, as we move down in the
hierarchy, the term embeddings learned based on the entire corpus have limited
power in capturing subtle semantics. For example, when splitting the machine
learning topic, the terms `machine learning' and `reinforcement learning' have
close global embeddings, and it is hard to discover quality sub-topics for the
machine learning topic.

\our consists of two modules for tackling the above challenges. The first is an
adaptive spherical clustering module for allocating terms to proper levels when
splitting a coarse topic. Relying on a ranking function that measures the
representativeness of different terms to each child topic, the clustering
module iteratively detects general terms that should remain in the parent topic
and keeps refining the clustering boundaries of the child topics.  The second
is a local term embedding module.  To enhance the discriminative power of term
embeddings at lower levels, \our employs an existing technique
\cite{Gui2017ExpertFI} that uses topic-relevant documents to learn local
embeddings for the terms in each topic. The local embeddings capture term
semantics at a finer granularity and are less constrained by the terms
irrelevant to the topic. As such, they are discriminative enough to separate
the terms with different semantics even at lower levels of the taxonomy.

We perform extensive experiments on two real data sets. Our qualitative results
show that \our can generate high-quality topic taxonomies, and our quantitative
analysis based on user study shows that \our outperforms baseline methods
significantly.

\section{Related Work}
\label{sect:related}

In this section, we review existing taxonomy construction methods, including
(1) pattern-based methods, (2) clustering-based methods, and (3) supervised
methods. 

\subsection{Pattern-Based Methods} 

A considerable number of pattern-based methods have been proposed to construct
hypernym-hyponym taxonomies wherein each node in the tree is an entity, and
each parent-child pair expresses the ``is-a'' relation. Typically, these works
first use pre-defined lexical patterns to extract hypernym-hyponym pairs from
the corpus, and then organize all the extracted pairs into a taxonomy tree.  In
pioneering studies, Hearst patterns like ``NP such as NP, NP, and NP'' were
proposed to automatically acquire hyponymy relations from text data
\cite{Hearst92}.  Then more kinds of lexical patterns have been manually
designed and used to extract relations from the web corpus
\cite{Seitner2016ALD, Panchenko2016TAXIAS} or Wikipedia
\cite{Ponzetto2007DerivingAL, Grefenstette2015INRIASACSH}.  With the
development of the Snowball framework, researchers teach machines how to
propagate knowledge among the massive text corpora using statistical approaches
\cite{agichtein2000snowball, zhu2009statsnowball}; Carlson et al. proposed a
learning architecture for Never-Ending Language Learning (NELL) in 2010
\cite{carlson2010toward}.  PATTY leveraged parsing structures to derive
relational patterns with semantic types and organizes the patterns into a
taxonomy \cite{nakashole2012patty}.  The recent MetaPAD \cite{jiang2017metapad}
used context-aware phrasal segmentation to generate quality patterns and group
synonymous patterns together for a large collection of facts.  Pattern-based
methods have demonstrated their effectiveness in finding particular relations
based on hand-crafted rules or generated patterns.  However, they are not
suitable for constructing a topic taxonomy because of two reasons.  First,
different from hypernym-hyponym taxonomies, each node in a topic taxonomy can
be a group of terms representing a conceptual topic. Second, pattern-based
methods often suffer from low recall due to the large variation of expressions
in natural language on parent-child relations.

\subsection{Clustering-Based Methods} 

A great number of clustering methods have been proposed for constructing
taxonomy from text corpus. These methods are more closely related to our
problem of constructing a topic taxonomy.  Generally, the clustering approaches
first learn the representation of words or terms and then organize them into a
structure based on their representation similarity \cite{BansalBMK14} and
cluster separation measures \cite{DaviesB79}.  Fu et al.  identified whether a
candidate word pair has hypernym-hyponym (``is-a'') relation by using the
word-embedding-based semantic projections between words and their hypernyms
\cite{FuGQCWL14}. Luu et al. proposed to use dynamic weighting neural network
to identify taxonomic relations via learning term embeddings \cite{LuuTHN16}.
Our method differs from these existing ones in two aspects. First, we do not
need labeled hypernym-hyponym pairs as supervision for learning either semantic
projections or dynamic weighting neural network.  Second, we employ a technique
called local embedding \cite{Gui2017ExpertFI} to learn embeddings for each
topic using only topic-relevant documents. The local embedding technique was first
proposed by Gui \etal \cite{Gui2017ExpertFI}. It captures fine-grained term
semantics with a local corpus and thus well separates terms with subtle
semantic differences.  On the term organizing end, Ciniano et al. used a
comparative measure to perform conceptual, divisive, and agglomerative
clustering for taxonomy learning \cite{cimiano2004comparing}.  Yang et al. also
used an ontology metric, a score indicating semantic distance, to induce
taxonomy \cite{YangC09}.  Liu et al.  used Bayesian rose tree to hierarchically
cluster a given set of keywords into a taxonomy \cite{liu2012automatic}.  Wang
et al.  adopted a recursive way to construct topic hierarchies by clustering
domain keyphrases \cite{Wang2013APM}.  Also, quite a number of hierarchical
topic models have been proposed for term organization
\cite{BleiGJT03,MimnoLM07,Downey2015EfficientMF}.  In our \our, we develop an
\textit{adaptive spherical clustering} module to allocate terms into proper
levels when we split a coarse topic. The module well groups terms of the same
topic together and separates child topics (as term clusters) with significant
distances.

\subsection{Supervised Methods} 

There have also been (semi-)supervised learning methods for taxonomy
construction \cite{kumar2001semi,KozarevaH10}.  Basically these methods extract
lexical features and learn a classifier that categorizes term pairs into
relations or non-relations, based on curated training data of hypernym-hyponym
pairs
\cite{YangC09,shearer2009exploiting,cui2010evolutionary,liu2012automatic}, or
syntactic contextual information harvested from NLP tools
\cite{wu2012probase,LuuKN14}. Recent techniques
\cite{Weeds2014LearningTD,Yu2015LearningTE,
  LuuTHN16,FuGQCWL14,Anke2016SupervisedDH} in this category leverage
pre-trained word embeddings and then use curated hypernymy relation datasets to
learn a relation classifier.  However, the training data for all these methods
are limited to extracting hypernym-hyponym relations and cannot be easily
adapted for constructing a topic taxonomy.  Furthermore, for massive
domain-specific text data (like scientific publication data we used in this
work), it is hardly possible to collect a rich set of supervised information
from experts.  Therefore, we focus on technical developments in
unsupervised taxonomy construction.

\section{Problem Description}
\label{sect:prelim}

The input for constructing a topic taxonomy includes two parts: (1) a corpus
$\D$ of documents; and (2) a set $\T$ of seed terms. The seed terms in $\T$ are
the key terms from $\D$, representing the terms of interest for taxonomy
construction\footnote{The term set can be either specified by end users or
  extracted from the corpus. In our experiments, we extract frequent noun
  phrases from $\D$ to form the term set $\T$.}.  Given the corpus $\D$ and the
term set $\T$, we aim to build a tree-structured hierarchy $\H$. Each node $C
\in \H$ denotes a conceptual topic, which is described by a set of terms $\T_C
\in \T$ that are semantically coherent.  Suppose a node $C$ has a set of
children $\S_C = \{S_1, S_2, \ldots, S_N\}$, then each $S_n (1 \le n \le N)$
should be a sub-topic of $C$, and have the same semantic granularity with its
siblings in $\S_C$.

\section{The \our Method}

In this section, we describe our proposed \our method. We first give an
overview of it in Section \ref{sect:overview}.  Then we introduce the details of
the adaptive spherical clustering and local embedding modules in Section
\ref{sect:cluster} and \ref{sect:local}, respectively.

\subsection{Method Overview}
\label{sect:overview}

\begin{figure*}[t]
  \centering
  \centerline{\includegraphics[width=0.8\textwidth]{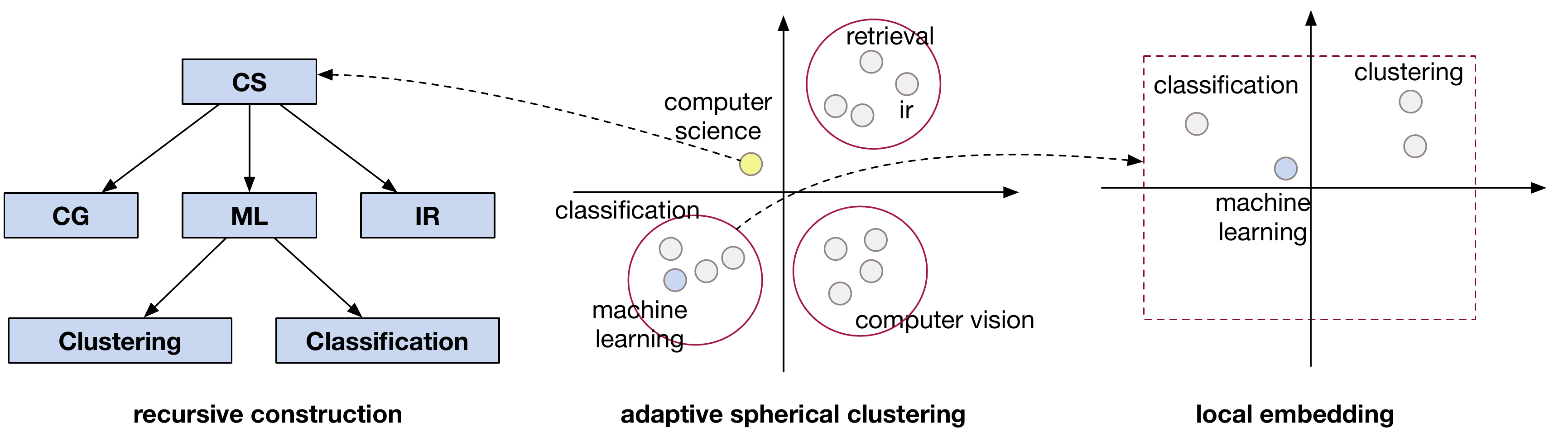}} 
  \caption{An overview of \our. It uses term embeddings to construct the
    taxonomy in a top-down manner, with two novel components for ensuring the
    quality of the resursive process: (1) an adaptive clustering module that
    allocates terms to proper topic nodes; and (2) a local embedding module for
    learning term embeddings on topic-relevant documents (courtesy of
    \cite{Gui2017ExpertFI}).
  }
  \label{fig:method}
\end{figure*}

In a nutshell, \our embeds all the concept terms into a latent space to capture
their semantics, and uses the term embeddings to build the taxonomy
recursively. As shown in Figure \ref{fig:method}, at the top level, we
initialize a root node containing all the terms from $\T$, which represents the
most general topic for the given corpus $\D$.  Starting from the root node, we
generate fine-grained topics level by level via top-down spherical clustering.
The top-down construction process continues until a maximum number of levels
$L_{max}$ is reached.

Given a topic $C$, we use spherical clustering to split $C$ into a set of
fine-grained topics $\S_C = \{S_1, S_2, \ldots, S_N\}$.  As mentioned earlier,
there are two challenges that need to be addressed in the resursive
construction process: (1) when splitting a topic $C$, it is problematic to
directly divide the terms in $C$ into sub-topics, because general terms should
remain in the parent topic $C$ instead of being allocated to any sub-topics;
(2) when we move down to lower levels, global term embeddings learned on the
entire corpus are inadequate for capturing subtle term semantics. In the
following, we introduce the adaptive clustering and local embedding modules in
\our for addressing these two challenges.

\subsection{Adaptive Spherical Clustering}
\label{sect:cluster}

The adaptive clustering module in \our is designed to split a coarse topic
$C$ into fine-grained ones. It is based on the spherical $K$-means algorithm
\cite{DhillonM01}, which groups a given set of term embeddings into $K$
clusters such that the terms in the same cluster have similar embedding
directions. Our choice of the spherical $K$-means algorithm is motivated by the
effectiveness of the cosine similarity \cite{mikolov2013distributed} in
quantifying the similarities between word embeddings. The center direction of a
topic acts as a semantic focus on the unit sphere, and the member terms of
that topic falls around the center direction to represent a coherent semantic
meaning.

\subsubsection{The adaptive clustering process.}

Given a coarse topic $C$, a straightforward idea for generating the
sub-topics of $C$ is to directly apply spherical K-means to $C$, such that
the terms in $C$ are grouped into $K$ clusters to form $C$'s sub-topics.
Nevertheless, such a straightforward strategy is problematic because not all
the terms in $C$ should be allocated into the child topics.  For example, in
Figure \ref{fig:method}, when splitting the root topic of computer science,
terms like `computer science' and `cs' are general --- they do not belong to
any specific child topics but instead should remain in the parent.
Furthermore, the existence of such general terms makes the clustering process
more challenging. As such general terms can co-occur with various contexts in
the corpus, their embeddings tend to fall on the boundaries of different
sub-topics. Thus, the clustering structure for the sub-topics is blurred,
making it harder to discover clear sub-topics.

\begin{algorithm}[t]
  \label{alg:clustering}
  \caption{Adaptive clustering for topic splitting.}
  \KwIn{
    A parent topic $C$; the number of sub-topics $K$; the term
    representativeness threshold $\delta$.
  }
  \KwOut{$K$ sub-topics of $C$.}
  $C_{sub} \gets C$\;
  \While{True} {
    $S_1, S_2, \ldots, S_K \gets \kmeans(C_{sub}, K)$\;
    \For{$k$ ~ from ~1 ~to ~$K$} {
      \For{$t \in S_k$} {
        $r(t, S_k) \gets $ representativeness of term $t$ for $S_k$\;
        \If{$r(t, S_k) < \delta$}{
          $S_k \gets S_k - \{t\}$\;
        }
      }
    }
    $C'_{sub} \gets S_1 \cup S_2 \cup \ldots \cup S_K$\;
    \If{$ C'_{sub} = C_{sub}$} {
      Break\;
    }
    $C_{sub} \gets C'_{sub}$\;
  }
  Return $S_1, S_2, \ldots, S_K$\;
\end{algorithm}

Motivated by the above, we propose \emph{an adaptive clustering module} in
\our.  As shown in Figure \ref{fig:method}, the key idea is to iteratively
identify general terms and refine the sub-topics after pushing general terms
back to the parent. Identifying general terms and refining child topics are two
operations that can mutually enhance each other: excluding the general terms in
the clustering process can make the boundaries of the sub-topics clearer; while
the refined sub-topics boundaries enable detecting additional general terms.

Algorithm 1 shows the process for adaptive spherical clustering. As shown,
given a parent topic $C$, it first puts all the terms of $C$ into the
sub-topic term set $C_{sub}$.  Then it iteratively identifies general terms
and refines the sub-topics.  In each iteration, it computes the
representativeness score of a term $t$ for the sub-topic $S_k$, and excludes
$t$ if its representativeness is smaller than a threshold $\delta$. After
pushing up general terms, it re-forms the sub-topic term set $C_{sub}$ and
prepares for the next spherical clustering operation. The iterative process
terminates when no more general terms can be detected, and the final set of
sub-topics $S_1, S_2, \ldots, S_K$ are returned.

\subsubsection{Measuring term representativeness.}

In Algorithm 1, the key question is how to measure the representativeness of a
term $t$ for a sub-topic $S_k$.  While it is tempting to measure the
representativeness of $t$ by its closeness to the center of $S_k$ in the
embedding space, we find such a strategy is unreliable: general terms may also
fall close to the cluster center of $S_k$, which renders the embedding-based
detector  inaccurate.

Our insight for addressing this problem is that, a representative term for
$S_k$ should appear frequently in $S_k$ but not in the sibling topics of
$S_k$. We hence measure term representativeness using the documents that belong
to $S_k$.  Based on the cluster memberships of terms, we first use the TF-IDF
scheme to obtain the documents belonging to each topic $S_k$. With these
$S_k$-related documents, we consider the following two factors for computing
the representativeness of a term $t$ for topic $S_k$:

\bitem{
\item
  \textbf{Popularity}: A representative term for $S_k$ should appear frequently
  in the documents of $S_k$.
\item
  \textbf{Concentration}: A representative term for $S_k$ should be much more
  relevant to $S_k$ compared to the sibling topics of $S_k$. 
}

To combine the above two factors, we notice that they should have conjunctive
conditions, namely a representative term should be both popular and
concentrated for $S_k$. Thus we define the representativeness of term $t$ for
topic $S_k$ as
\begin{equation}
  \label{eqn:rank}
  r(t, S_k) = \sqrt{pop(t, S_k) \cdot con(t, S_k)}
\end{equation}
where $pop(t, S_k)$ and $con(t, S_k)$ are the popularity and concentration
scores of $t$ for $S_k$.  Let $\D_k$ denotes the documents belonging to  $S_k$,
we define $pop(t, S_k)$ as the normalized frequency of $t$ in $\D_k$:
$$
pop(t, S_k) = \frac{\log (tf(t, \D_k) + 1)}{\log tf(\D_k)},
$$
where $tf(t, \D_k)$ is number of occurrences of term $t$ in $\D_k$, and
$tf(\D_k)$ is the total number of tokens in $\D_k$.

To compute the concentration score, we first form a pseudo document $D_k$ for
each sub-topic $S_k$ by concatenating all the documents in $\D_k$. Then we
define the concentration of term $t$ on $S_k$ based on its relevance to the
pseudo document $D_k$:
$$
con(t, S_k) = \frac{\exp( rel(t, D_k) )}{1 + \Sum_{1 \le j \le K} \exp( rel(t,
  D_j))},
$$
where $rel(p, D_k)$ is the BM25 relevance of term $t$ to the pseudo document
$D_k$.


\begin{example}
  Figure \ref{fig:method} shows the adaptive clustering process for splitting
  the computer science topic into three sub-topics: computer graphics (CG),
  machine learning (ML), and information retrieval (IR). Given a sub-topic, for
  example ML, terms (\eg, `clustering', `classificiation') that are popular and
  concentrated in this cluster receive high representativeness scores.  In
  contrast, terms (\eg, `computer science') that are not representative for any
  sub-topics are considered as general terms and pushed back to the parent.
\end{example}

\subsection{Local Embedding \cite{Gui2017ExpertFI}}
\label{sect:local}

The recursive taxonomy construction process of \our relies on term embeddings,
which encode term semantics by learning fixed-size vector representations for
the terms. We use the SkipGram model \cite{mikolov2013distributed} for learning
term embeddings.  Given a corpus, SkipGram models the relationship between a
term and its context terms in a sliding window, such that the terms that share
similar contexts tend to have close embeddings in the latent space.  The result
embeddings can well capture the semantics of different terms and been
demonstrated useful for various NLP tasks.

Formally, given a corpus $\D$, for any token $t$, we consider a sliding window
centered at $t$ and use $W_t$ to denote the tokens appearing in the context
window. Then we define the log-probability of observing the contextual terms as 
$$
\log p(W_t | t) = \Sum_{w \in W_t} \log p(w | t) = 
\Sum_{w \in W_t} \log \frac{\v_t \v'_w}{\Sum_{w' \in V} \v_t \v'_{w'}}
$$
where $\v_t$ is the embedding for term $t$,  $\v'_w$ is the contextual
embedding for the term $w$, and $V$ is the vocabulary of the corpus $\D$.
Then the overall objective function of SkipGram is defined over all 
the tokens in $\D$, namely
$$
L = \Sum_{t \in \D} \Sum_{w \in W_t} \log p(w | t),
$$
and the term embeddings can be learned by maximizing the objective with
stochastic gradient descent and negative sampling
\cite{mikolov2013distributed}.

However, when we use the term embeddings trained on the entire corpus $\D$ for
taxonomy construction, one drawback is that these global embeddings have
limited discriminative power at lower levels.  Let us consider the term
`reinforcement learning' in Figure \ref{fig:method}. In the entire corpus $\D$, it
shares a lot of  similar contexts with the term `machine learning', and thus
has an embedding close to `machine learning' in the latent space. The
proximity with `machine learning' makes it successfully assigned into the
machine learning topic when we are splitting the root topic.  Nevertheless,
as we move down to split the machine learning topic, the embeddings of
`reinforcement learning' and other machine learning terms are entangled together,
making it difficult to discover sub-topics for machine learning.

As introduced in \cite{Gui2017ExpertFI}, local embedding is able to capture the
semantic information of terms at finer granularity. Therefore, we employ it to
enhance the discriminative power of term embeddings at lower levels of the
taxonomy.  Here we describe how to use it for obtaining discriminative
embeddings for the taxonomy construction task.  For any topic $C$ that is not
the root topic, we learn local term embeddings for splitting $C$.
Specifically, we first create a sub-corpus $\D_{C}$ from $\D$ that is relevant
to the topic $C$.  To obtain the sub-corpus $\D_{C}$, we employ the following
two strategies: (1) \emph{Clustering-based.} We derive the cluster membership
of each document $d \in \D$ by aggregating the cluster memberships of the terms
in $d$ using TF-IDF weight. The documents that are clustered into topic $C$ are
collected to form the sub-corpus $\D_C$.  (2) \emph{Retrieval-based.} We
compute the embedding of any document $d \in \D$ using TF-IDF weighted average
of the term embeddings in $d$. Based on the obtained document embeddings, we
use the mean direction of the topic $C$ as a query vector to retrieve the
top-$M$ closest documents and form the sub-corpus $\D_C$. In practice, we use
the first strategy as the main one to obtain $\D_C$, and apply the second
strategy for expansion if the clustering-based subcorpus is not large enough.
Once the sub-corpus $\D_C$ is retrieved, we apply the SkipGram model to the
sub-corpus $\D_{C}$ to obtain term embeddings that are tailored for splitting
the topic $C$.

\begin{example}
Consider Figure \ref{fig:method} as an example, when splitting the machine
learning topic, we first obtain a sub-corpus $\D_{ml}$ that is relevant to
machine learning. Within $\D_{ml}$, terms reflecting general machine learning
topics such as `machine learning' and `ml' appear in a large number of
documents. They become similar to stopwords and can be easily separated from
more specific terms.  Meanwhile, for those terms that reflect different machine
learning sub-topics (\eg, `classifcation' and `clustering'), they are also
better separated in the local embedding space.  Since the local embeddings are
trained to preserve the semantic information for topic-related documents,
different terms have more freedom to span in the embedding space to reflect
their subtle semantic differences.
\end{example}

\section{Experiments}
\label{sect:exp}

\subsection{Experimental Setup}

\subsubsection{Datasets}

We use two datasets in our experiments: (1) \dblp contains around 1,889,656
titles of computer science papers from the areas of information retrieval,
computer vision, robotics, security \& network, and machine learning. From
those paper titles, we use an existing NP chunker to extract all the noun
phrases and then remove infrequent ones to form the term set, resulting in
13,345 distinct terms; (2) \sigp contains 94,476 paper abstracts from the area
of signal processing.  Similarly, we extract all the noun phrases in those
abstracts to form the term set and obtain 6,982 different terms.\footnote{The
  code and data are available at
  https://github.com/franticnerd/taxogen/.}

\subsubsection{Compared Methods}

We compare \our with the following baseline methods that are capable of
generating topic taxonomies:

\benum{

\item \hlda (hierarchical Latent Dirichlet Allocation) \cite{BleiGJT03} is a
  non-parametric hierarchical topic model. It models the probability of
  generating a document as choosing a path from the root to a leaf and sampling
  words along the path.  We apply \hlda for topic-level taxonomy construction
  by regarding each topic in \hlda as a topic.

\item \hpam (hierarchical Pachinko Allocation Model) is a state-of-the-art
  hierarchical topic model \cite{MimnoLM07}.  Different from \our that
  generates the taxonomy recursively, \hpam takes all the documents as
  its input and outputs a pre-defined number of topics at different levels
  based on the Pachinko Allocation Model.

\item \hc (hierarchical clustering) uses hierarchical clustering for taxonomy
  construction. We first apply the SkipGram model on the entire corpus to learn
  term embeddings, and then use spherical k-means to cluster those embeddings 
  in a top-down manner.

\item \noac is a variant of \our without the adaptive clustering module.  In
  other words, when splitting one coarse topic into fine-grained ones, it
  simply performs spherical clustering to group parent terms into child
  topics.

\item \nole is a variant of \our without the local embedding module. During the
  recursive construction process, it uses the global embeddings that are
  learned on the entire corpus throughout the construction process.

}

\subsubsection{Parameter Settings}

We use the methods to generate a four-level taxonomy on \dblp and a three-level
taxonomy on \sigp. There are two key parameters in \our: the number $K$ for
splitting a coarse topic and the representativeness threshold $\delta$ for
identifying general terms.  We set $K = 5$ as we found such a setting matches
the intrinsic taxonomy structures well on both \dblp and \sigp.  For $\delta$,
we set it to $0.25$ on \dblp and $0.15$ on \sigp after tuning, because we
observed such a setting can robustly detect general terms that belong to parent
topics at different levels in the construction process.  For \hlda, it
involves three hyper-parameters: (1) the smoothing parameter $\alpha$ over
level distributions; (2) the smoothing parameter $\gamma$ for the Chinese
Restaurant Process; and (3) the smoothing parameter $\eta$ over topic-word
distributions.  We set $\alpha = 0.1, \gamma = 1.0, \eta = 1.0$. Under such a
setting, \hlda generates a comparable number of topics with \our on both
datasets. The method \hpam requires to set the mixture priors for super- and
sub-topics.  We find that the best values for these two priors are 1.5 and 1.0
on \dblp and \sigp, respectively.  The remaining three methods (\hc, \noac, and
\nole) have a subset of the parameters of \our, and we set them to the same
values as \our.

\subsection{Qualitative Results}

In this subsection, we demonstrate the topic taxonomies generated by
different methods on \dblp. We apply each method to generate a four-level
taxonomy on \dblp, and each parent topic is split into five child
topics by default (except for \hlda, which automatically determines the
number of child topics based on the Chinese Restaurant Process).

\begin{figure*}[h]
  \centering {\
    \subfigure[The sub-topics generated by \our under the topics `*' (level 1),
    `information retrieval' (level 2), and `Web search' (level 3).] {
      \label{fig:tax-ir}
      \includegraphics[width=0.9\textwidth]{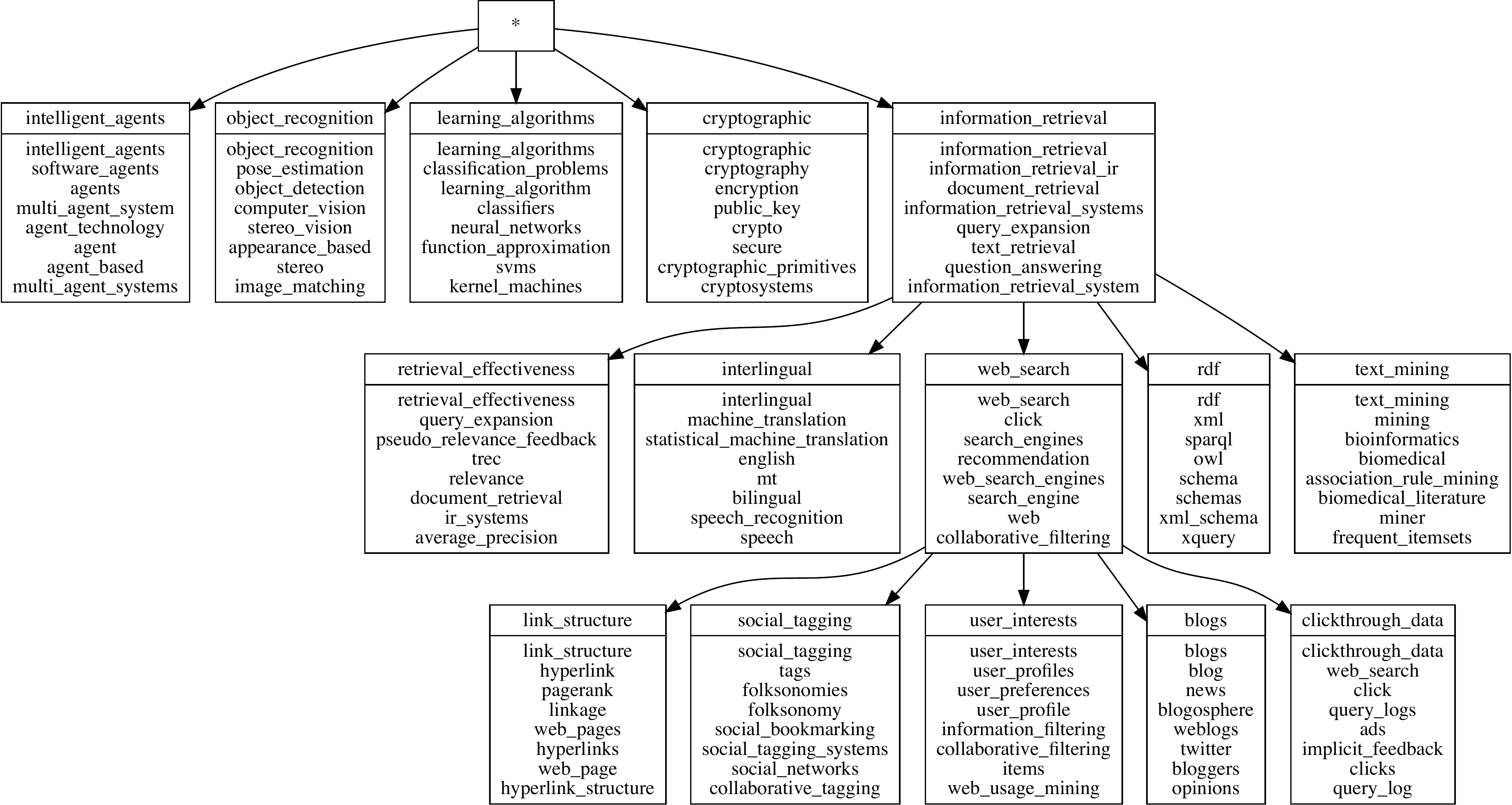}
    }
    \hfil
    \subfigure[The sub-topics generated by \our under the topics
    `learning algorithms' (level 2) and  `neural network' (level 3).] {
      \label{fig:tax-ml}
      \includegraphics[width=1.0\textwidth]{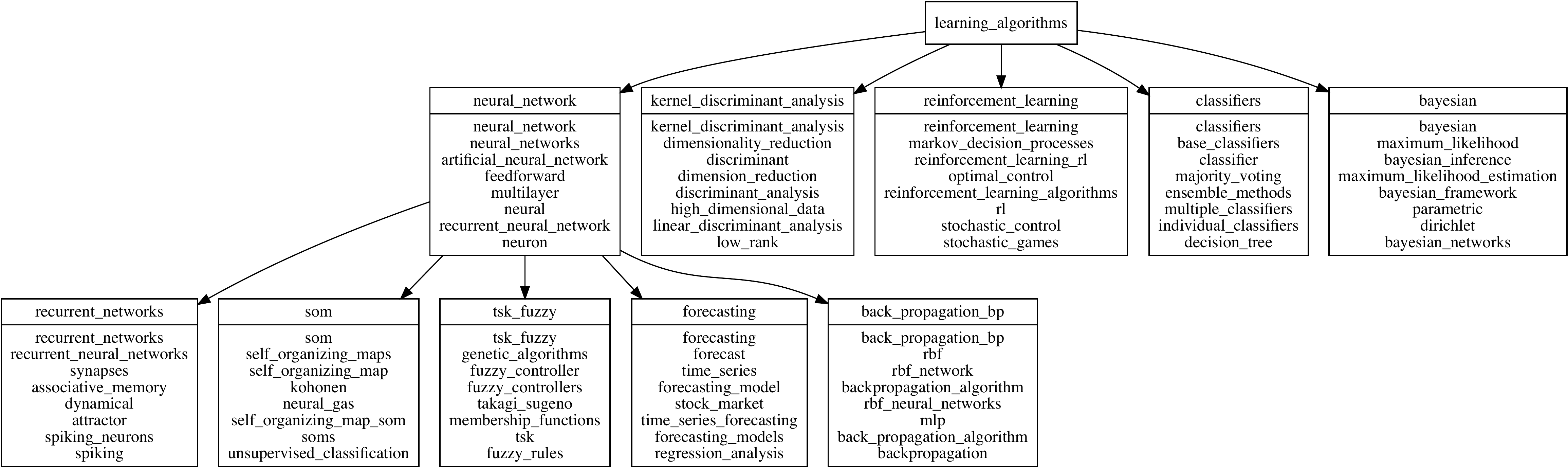}
    }
  }
  \vspace{-3ex}
  \caption{Parts of the taxonomy generated by \our on the \dblp dataset. For
    each topic, we show its label and the top-eight representative terms
    generated by the ranking function of \our.  All the labels and terms are
    returned by \our automatically without manual selection or filtering.}
  \label{fig:our-tax}
\end{figure*}

Figure \ref{fig:our-tax} shows parts of the taxonomy generated by \our. As
shown in Figure \ref{fig:tax-ir}, given the \dblp corpus, \our splits the root
topic into five sub-topics: `intelligent agents', `object recognition',
`learning algorithms', `cryptographic', and `information retrieval'. The labels
for those topics are generated automatically by selecting the term that is
most representative for a topic (Equation \ref{eqn:rank}). We find those
labels are of good quality and precisely summarize the major research areas
covered by the \dblp corpus.  The only minor flaw for the five labels is
`object recognition', which is too specific for the computer vision area. The
reason is probably because the term `object recognition' is too popular in the
titles of computer vision papers, thus attracting the center of the spherical
cluster towards itself.

\begin{figure*}[h]
  \centering {\
    \subfigure[The sub-topics generated by \nole under the topic `web search' (level 3).] {
      \label{fig:nole-ws}
      \includegraphics[height=0.15\textwidth]{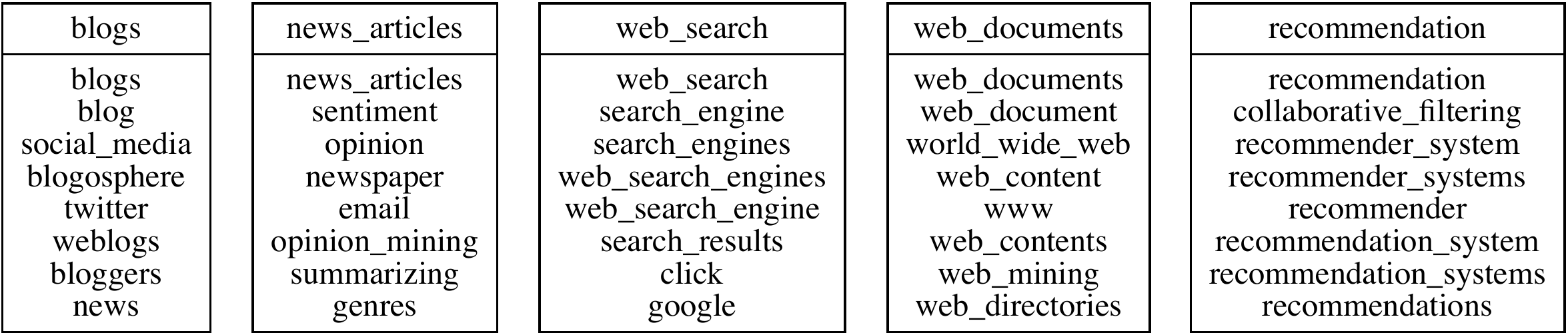} 
    }
    \hfil
    \subfigure[The sub-topics generated by \nole under the topic `neural networks' (level 3).] {
      \label{fig:nole-nn}
      \includegraphics[height=0.15\textwidth]{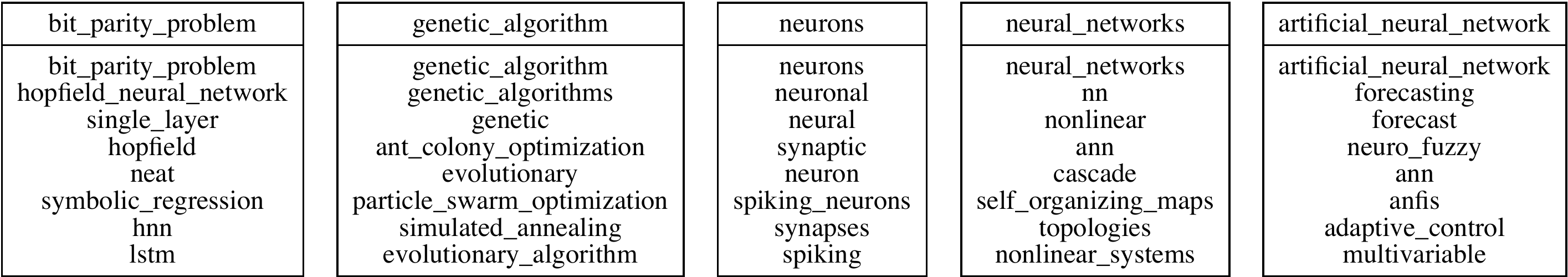}
    }
    \hfil
    \subfigure[The sub-topics generated by \noac under the topic `neural network' (level 3).] {
      \label{fig:noac-nn}
      \includegraphics[height=0.15\textwidth]{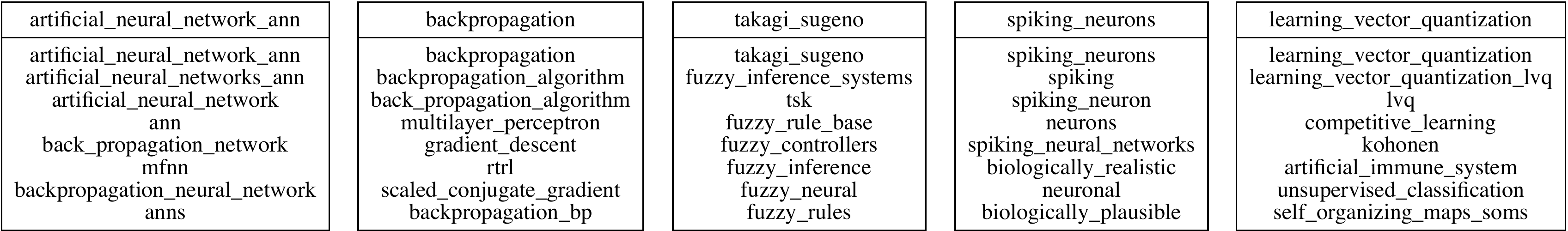}
    }
  }
  \vspace{-3ex}
  \caption{Example topics generated by \nole and \noac on the \dblp
    dataset. Again,  we show the label and the top-eight representative
    terms for each topic.}
  \label{}
\end{figure*}

In Figure \ref{fig:tax-ir} and \ref{fig:tax-ml}, we also show how \our splits
level-two topics `information retrieval' and `learning algorithms' into more
fine-grained topics. Taking `information retrieval' as an example: (1) at
level three, \our can successfully find major areas in information retrieval:
retrieval effectiveness, interlingual, Web search, rdf \& xml query, and text
mining; (2) at level four, \our splits the Web search topic into more
fine-grained problems: link analysis, social tagging, recommender systems \&
user profiling, blog search, and clickthrough models. Similarly for the machine
learning topic (Figure \ref{fig:tax-ml}), \our can discover level-three
topics like `neural network' and level-four topic like `recurrent neural
network'. Moreover, the top terms for each topic are of good quality --- they
are semantically coherent and cover different aspects and expressions of the
same topic.

We have also compared the taxonomies generated by \our and other baseline
methods, and found that \our offers clearly better taxonomies from the
qualitative perspective.  Due to the space limit, we only show parts of the
taxonomies generated by \noac and \nole to demonstrate the effectiveness of
\our.  As shown in Figure \ref{fig:nole-ws}, \nole can also find several
sensible child topics for the parent topic (\eg, `blogs' and `recommender
system' under `Web search'), but the major disadvantage is that a considerable
number of the child topics are false positives. Specifically, a number of
parent-child pairs (`web search' and `web search', `neural networks' and
`neural networks') actually represent the same topic instead of true
hypernym-hyponym relations.  The reason behind is that \nole uses global term
embeddings at all levels, and thus the terms for different semantic
granularities have close embeddings and hard to be separated at lower levels.
Such a problem also exists for \noac, but with a different reason: \noac does
not leverage adaptive clustering to push up the terms that belong to the parent
topic. Consequently, at fine-grained levels, terms that have different
granularities are all involved in the clustering step, making the clustering
boundaries less clear compared to \our.  Such qualitative results clearly show
the advantages of \our over the baseline methods, which are the key factors
that leads to the performance gaps between them in our quantitative evaluation.

Table \ref{tab:local} further compares global and local term embeddings for
similarity search tasks. As shown, for the given two queries, the top-five
terms retrieved with global embeddings (\ie, the embeddings trained on the
entire corpus) are relevant to the queries, yet they are semantically
dissimilar if we inspect them at a finer granularity. For example, for the
query `information extraction', the top-five similar terms cover various areas
and semantic granularities in the NLP area, such as `text mining', `named
entity recognition', and `natural language processing'. In contrast, the
results returned based on local embeddings are more coherent and of the same
semantic granularity as the given query.

\begin{table}[t]
\centering
\caption{Similarity searches on \dblp for: (1) Q1 = `pose\_estimation'; and (2)
  Q2 = `information\_extraction'. For both queries, we use cosine similarity to
  retrieve the top-five terms in the vocabulary based on global and local
  embeddings.  The local embedding results for `pose\_estimation' are obtained
  in the `object\_recognition' sub-topic, while the results for
  `information\_extraction' are obtained in the `learning\_algorithms'
  sub-topic.
}
\label{tab:local}
\begin{small}
\begin{tabular}{@{}ccc@{}}
\toprule
\textbf{Query}                           & \textbf{Global Embedding}     & \textbf{Local Embedding}      \\ \midrule
\multirow{5}{*}{Q1}        & pose\_estimation              & pose\_estimation              \\
                                         & single\_camera                & camera\_pose\_estimation      \\
                                         & monocular                     & dof                           \\
                                         & d\_reconstruction             & dof\_pose\_estimation         \\
                                         & visual\_servoing              & uncalibrated                  \\\midrule
                                     \multirow{5}{*}{Q2} & information\_extraction       & information\_extraction       \\
                                         & information\_extraction\_ie   & information\_extraction\_ie   \\
                                         & text\_mining                  & ie                            \\
                                         & named\_entity\_recognition    & extracting\_information\_from \\
                                         & natural\_language\_processing & question\_anwering\_qa        \\ \bottomrule
\end{tabular}
\end{small}
\end{table}

\subsection{Quantitative Analysis}

In this subsection, we quantitatively evaluate the quality of the constructed
topic taxonomies by different methods. The evaluation of a taxonomy is a
challenging task, not only because there are no ground-truth taxonomies for our
used datasets, but also that the quality of a taxonomy should be judged from
different aspects. In our study, we consider the following aspects for
evaluating a topic-level taxonomy:

\bitem{
\item
  \textbf{Relation Accuracy} aims at measuring the portions of the true
  positive parent-child relations in a given taxonomy.
\item
  \textbf{Term Coherency} aims at quantifying how semantically coherent the top
  terms are for a topic.
\item
  \textbf{Cluster Quality} examines whether a topic and its siblings form 
  quality clustering structures that are well separated in the semantic space.
}


We instantiate the evaluations of the above three aspects as follows. First,
for the relation accuracy measure, we take all the parent-child pairs in a
taxonomy and perform user study to judge these pairs. Specifically, we
recruited 10 doctoral and post-doctoral researchers in Computer Science as
human evaluators. For each parent-child pair, we show the parent and child
topics (in the form of top-five representative terms) to at least three
evaluators, and ask whether the given pair is a valid parent-child relation.
After collecting the answers from the evaluators, we simply use majority voting
to label the pairs and compute the ratio of true positives.  Second, to measure
term coherency, we perform a term intrusion user study.  Given the top five
terms for a topic, we inject into these terms a fake term that is randomly
chosen from a sibling topic.  Subsequently, we show these six terms to an
evaluator and ask which one is the injected term.  Intuitively, the more
coherent the top terms are, the more likely an evaluator can correctly identify
the injected term, and thus we compute the ratio of correct instances as the
term coherency score.  Finally, to quantify cluster quality, we use the
Davies-Bouldin (DB) Index measure:  For any cluster $C$, we first compute the
similarities between $C$ and other clusters and assign the largest value to $C$
as its cluster similarity.  Then the DB index is obtained by averaging all the
cluster similarities \cite{DaviesB79}. The smaller the DB index is, the better
the clustering result is.

\begin{table}[h]
\centering
\caption{Relation accuracy and  term coherency of different methods on the DBLP
  and SP datasets.}
\label{tab:score}
\begin{tabular}{@{}lcccc@{}}
\toprule
& \multicolumn{2}{c}{Relation Accuracy} & \multicolumn{2}{c}{Term Coherency} \\ \midrule
Method & DBLP                & SP                 & DBLP                & SP                 \\
\hpam  & 0.109               & 0.160              & 0.173               & 0.163              \\
\hlda  & 0.272               & 0.383              & 0.442               & 0.265              \\
\hc    & 0.436               & 0.240              & 0.467               & 0.571              \\
\noac  & 0.563               & 0.208              & 0.35                & 0.428               \\
\nole  & 0.645               & 0.240              & 0.704               & 0.510              \\
\our   & \textbf{0.775}      & \textbf{0.520}     & \textbf{0.728}      & \textbf{0.592}     \\ \bottomrule
\end{tabular}
\end{table}

Table \ref{tab:score} shows the relation accuracy and term coherency of
different methods. As shown, \our achieves the best performance in terms of
both measures.  \our significantly outperforms topic modeling methods as well
as other embedding-based baseline methods. Comparing the performance of \our,
\noac, and \nole, we can see both the adaptive clustering and the local
embedding modules play an important role in improving the quality of the result
taxonomy: the adaptive clustering module can correctly push background terms
back to parent topics; while the local embedding strategy can better capture
subtle semantic differences of terms at lower levels. For both measures, the
topic modeling methods (\hlda and \hpam) perform significantly worse than
embedding-based methods, especially on the short-document dataset \dblp. The
reason is two-fold. First, \hlda and \hpam make stronger assumptions on
document-topic and topic-term distributions, which may not fit the empirical
data well.  Second, the representative terms of topic modeling methods are
selected purely based on the learned multinomial distributions, whereas
embedding-based methods perform distinctness analysis to select terms that are
more representative.

\begin{figure}[h]
  \centering {\
    \subfigure[DB index on \dblp.] {
      \includegraphics[width=0.44\columnwidth]{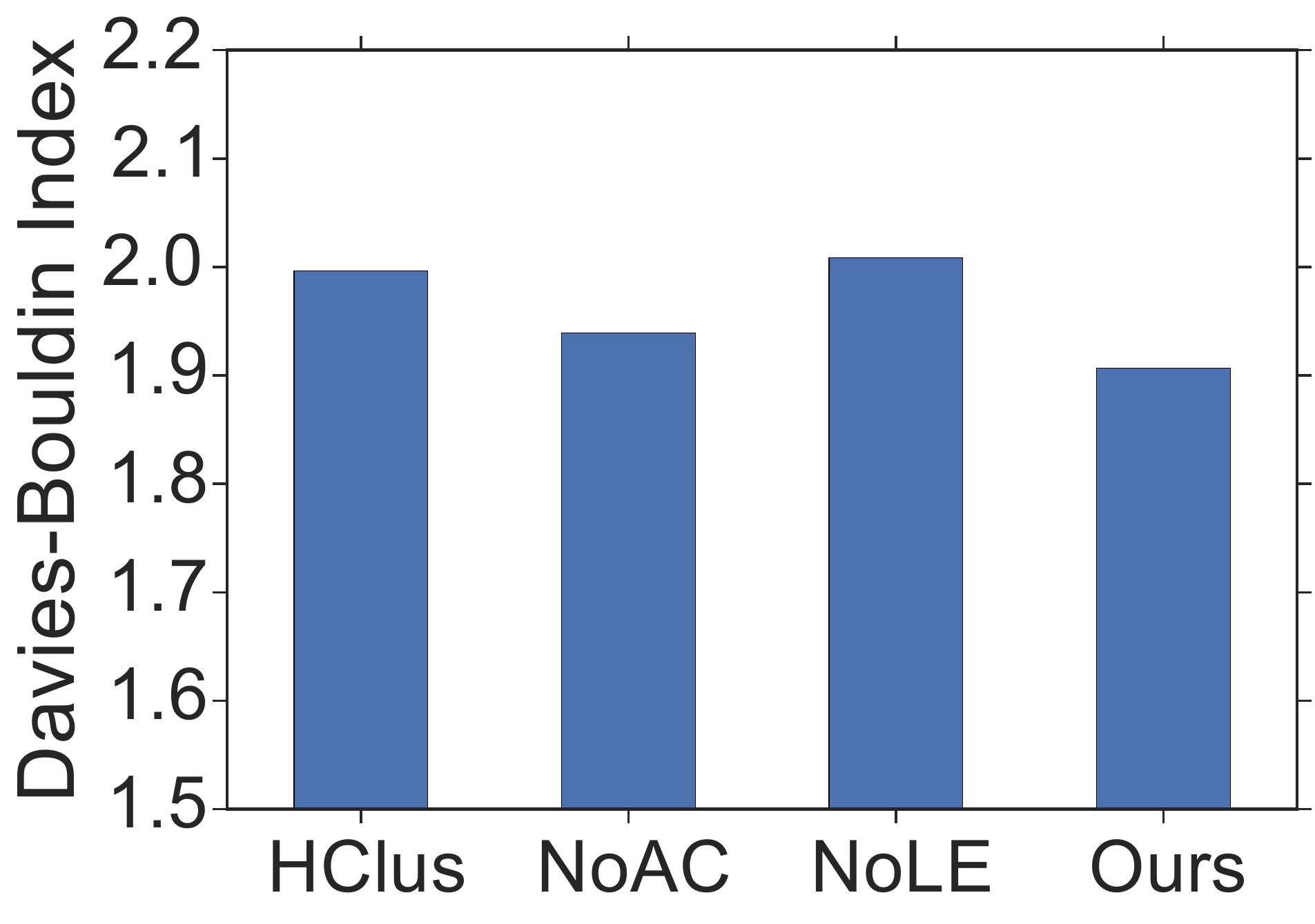}
    }
    \hfil
    \subfigure[DB index on \sigp.] {
      \includegraphics[width=0.44\columnwidth]{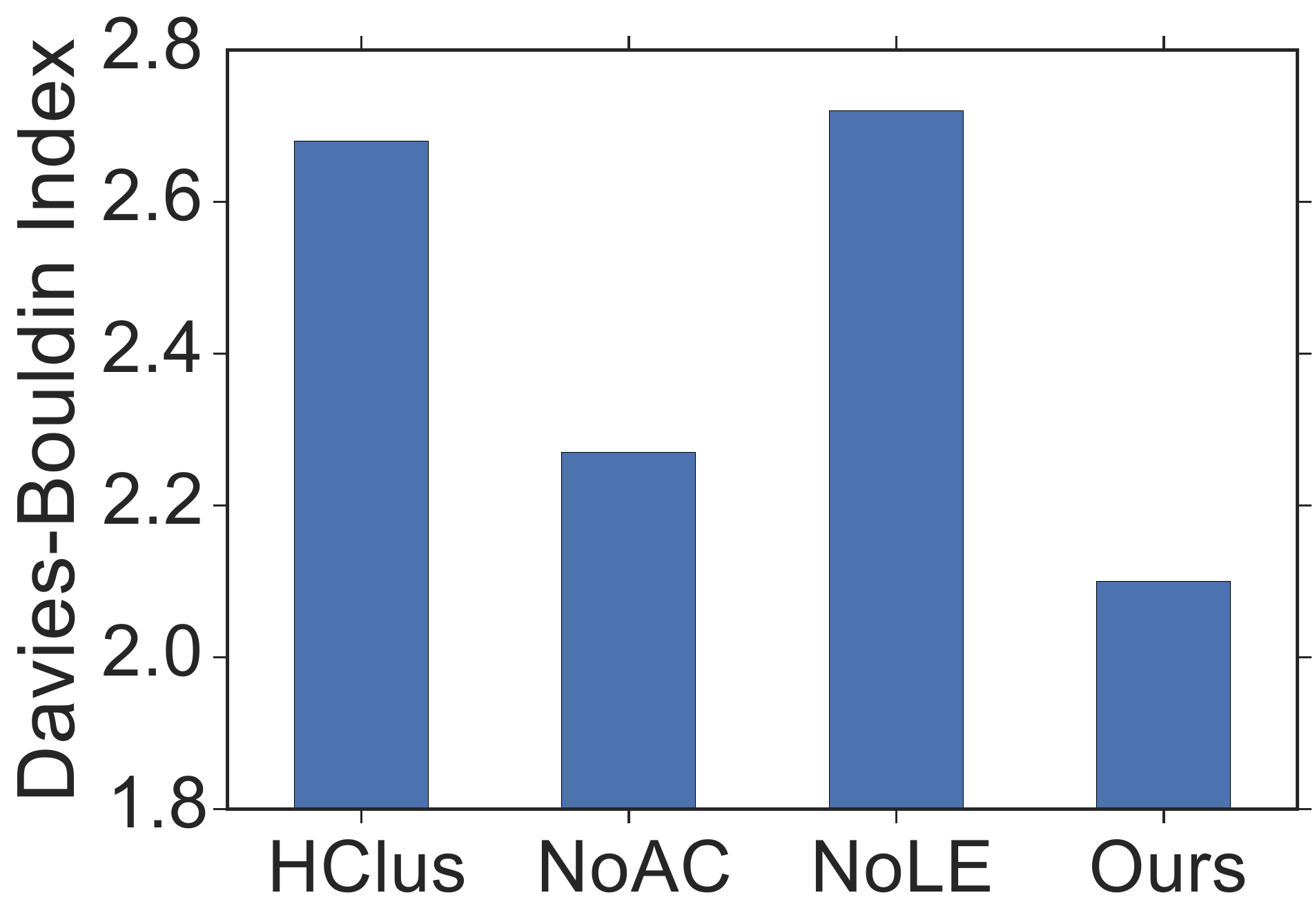}
    }
  }
  \vspace{-3ex}
  \caption{The Davies-Bouldin index of  embedding-based methods on 
    \dblp and \sigp.}
  \label{fig:cluster}
\end{figure}

Figure \ref{fig:cluster} shows the DB index of all the embedding-based methods.
\our achieves the smallest DB index (the best clustering result) among these
four methods. Such a phenomenon further validates the fact that both the
adaptive clustering and local embedding modules are useful in producing clearer
clustering structures: (1) The adaptive clustering process gradually identifies
and eliminates the general terms, which typically lie in the boundaries of
different clusters; (2) The local embedding module is capable of refining term
embeddings using a topic-constrained sub-corpus, allowing the sub-topics to
be well separated from each other at a finer granularity.

\section{Conclusion and Discussion}
\label{sect:conclusion}

We studied the problem of constructing topic taxonomies from a given text
corpus. Our proposed method \our relies on term embedding and spherical
clustering to construct a topic taxonomy in a recursive way. It consists of an
adaptive clustering module that allocates terms to proper levels when splitting
a coarse topic, as well as a local embedding module that learns term embeddings
to maintain strong discriminative power at lower levels.  In our experiments,
we have demonstrated that both two modules are useful in improving the quality
of the resultant taxonomy, which renders \our advantages over existing methods
for building topic taxonomies.  One limitation of the current version of \our
is that it requires a pre-specified number of clusters when splitting a coarse
topic into fine-grained ones.  In the future, it is interesting to extend \our
to allow it to automatically determine the optimal number of children for each
parent topic in the construction process. 

\vspace{2ex} \noindent\textbf{Acknowledgements:} This work was sponsored in
part by U.S. Army Research Lab. under Cooperative Agreement No.
W911NF-09-2-0053 (NSCTA), DARPA under Agreement No. W911NF-17-C-0099, National
Science Foundation IIS 16-18481, IIS 17-04532, and IIS-17-41317, DTRA
HDTRA11810026, and grant 1U54GM114838 awarded by NIGMS through funds provided
by the trans-NIH Big Data to Knowledge (BD2K) initiative (www.bd2k.nih.gov).
Any opinions, findings, and conclusions or recommendations expressed in this
document are those of the author(s) and should not be interpreted as the views
of any funding agencies.

\bibliographystyle{abbrv}
\bibliography{cited}

\end{document}